# Polarization amplitude-phase direction finding methods in two-canal UHF radio beacon navigation systems

V L Gulko[1], A A Mescheryakov[2]

[1]Candidate of Engineering Sciences, Associate Professor, Researcher of the Institute of Radio-engineering systems, TUSUR, Tomsk, Russia

[2]Candidate of Engineering Sciences, Associate Professor, Lead researcher of the Institute of Radio-engineering systems, TUSUR, Tomsk, Russia

E-mail: msch@rts.tusur.ru

There are investigated amplitude-phase method of the moving object bearing when there are used orthogonal linear polarized radio signals illuminated simultaneously from two horizontally spaced points with known co-ordinates. The bearing is measured onboard the moving object with two canal UHF system utilizing the amplitude-phase processing of the signals received within the linear polarization basis. Analysis is done for the general case when the orts angle of the linear polarization basis to the measuring plate is $\theta$.

The existing methods of amplitude-phase bearing by radio beacon navigation systems in order to find the moving object (MO) angle co-ordinates utilize amplitude, phase and time characteristics of radio beacon signals [1-3]. Polarization characteristics of the signals are not used as the source of the navigation information [4,5]. Polarization methods of the MO bearing were investigated in [6,8], in which orthogonally linearly polarized (OLP) radio beacon signals were used. In this case, the parameters of the linear polarized basis (LPB) were considered the same though it is known that theoretically there are many polarization basis's which can be used in polarization radio beacon navigation systems.

This publication is intended to study the amplitude-phase methods of direction finding of a MO using orthogonally linearly polarized radio beacon signals for cases when the unit orts of the LPB, in which the bearing is estimated, are generally oriented at an angle $\theta$ to the measurement plane.

**Analysis of the radar canal using Johns matrixes and vectors**

Let's suggest that the signals with equal amplitudes, initial phases and waves length are illuminated simultaneously from two points horizontally spaced by the distance d. The flat wave can be described by the Johns vector [4,6,9]. Then the OLP waves emitted by the radio beacon in the direction $\alpha$ can be represented by the Jones vector of the resulting wave in the LPB, omitting the time dependence, in the form [7]:

$$\vec{E}_p = \frac{1}{\sqrt{2}} \left\{ \begin{bmatrix} 1 \\ 0 \end{bmatrix} + \begin{bmatrix} 0 \\ 1 \end{bmatrix} \cdot e^{j\Delta\varphi} \right\}, \quad (1)$$

where $\Delta\varphi = \frac{2\pi d}{\lambda} \sin\alpha$ is the phase difference between the OLP waves at the receiving point on a MO.

Equation (1) allows to find the bearing as the angle between perpendicular at the base center $d$ and the MO direction [7]:

$$\alpha = \pm \arcsin \frac{\lambda}{2\pi d} \cdot (\Delta\varphi) \pm n\pi \quad (2)$$

where $n = 0, 1, 2...$

After transformation (1) becomes [7,10]:

$$\vec{E}_p = \frac{1}{\sqrt{2}} \left\{ \begin{matrix} 1 \\ e^{j\Delta\varphi} \end{matrix} \right\}. \quad (3)$$

The presence of the factor $1/\sqrt{2}$ in (3) is due to the unit intensity of the resulting wave adopted for convenience. Superposition of the OLP wave with the same amplitudes but different phases (1) in general case provides elliptic polarization of the resulting wave. The angle $\beta$ of the polarization ellipse can have two values [10]:

$$\beta_1 = \pi/4 \text{ by } 0 \le \Delta\varphi < \pi/2, \quad (4)$$

$$\beta_2 = 3\pi/4 \text{ by } \pi/2 < \Delta\varphi \le \pi. \quad (5)$$

The ellipticity coefficient module $|r|$ changes [10]:

$$|r| = (0 \div 1) \text{ by } 0 \le \Delta\varphi \le \pi/2, \quad (6)$$

and

$$|r| = (1 \div 0) \text{ by } \pi/2 \le \Delta\varphi \le \pi. \quad (7)$$

When $\Delta\varphi = \pm\pi/2$ (there is passed the round condition of the resulting wave polarization) the angle of the polarization ellipse jumps from $\beta_1 = \pi/4$ to $\beta_2 = 3\pi/4$ and backward. When the MO position corresponds the phase shift $\Delta\varphi = 0$ or $\Delta\varphi = \pi$, the resulting wave is linearly polarized with the electric field vector orientation $\beta_1 = \pi/4$ or $\beta_2 = 3\pi/4$ [10].

So, polarization characteristics of the resulting wave (3) depends on the direction of illumination and in general case polarization of the resulting wave is elliptic while due to the orthogonality of the illuminated waves its intensity does not depend on the direction of illumination.

Suppose that the reception of the resultant wave (3) on board the MO is carried out by a receiving antenna, in the microwave path of which a linear polarization separator (LPS) is installed. Orts of the LPS co-ordinate system coincide with the axes of the orthogonal put right angled wave guides [4,7,8] and also coincide with the LPB orts oriented at the angle $\theta$ to the plate of measurement. The chosen orts orientation allows to divide the received wave (3) in two orthogonal linearly polarizes components $\vec{E}_1$ and $\vec{E}_2$ oriented along the LPB orts vector. The Jones vectors $\vec{E}_1$ and $\vec{E}_2$ at the outputs of the LPB arms, oriented generally at an angle $\theta$, can be found (omitting the time dependence) as a result of the transformation:

$$\vec{E}_1 = [\Pi_1][R(\theta)]\vec{E}_p; \qquad (8)$$

$$\vec{E}_2 = [\Pi_2][R(\theta)]\vec{E}_p \qquad (9)$$

where $\vec{E}_p$ is the Johns vector of the resulting wave (3),

$[R(\theta)] = \begin{bmatrix} \cos\theta & \sin\theta \\ -\sin\theta & \cos\theta \end{bmatrix}$ - the counterclockwise rotation operator by the angle $\theta$ [9];

$[\Pi_1] = \begin{bmatrix} 0 & 0 \\ 0 & 1 \end{bmatrix}$ - Johns operator of the first LPS shoulder (transition from the round to the right-angled wave guide with own vertical polarization) [9];

$[\Pi_2] = \begin{bmatrix} 1 & 0 \\ 0 & 0 \end{bmatrix}$ - Johns operator of the second LPS shoulder (transition from the round to the right-angled wave guide with own horizontal polarization) [9];

After calculation there are two Johns vector $\vec{E}_1$ and $\vec{E}_2$ at the LPS outputs:

$$\vec{E}_1(\theta) = \begin{bmatrix} 0 \\ -\sin\theta + \cos\theta e^{j\Delta\varphi} \end{bmatrix}; \qquad (10)$$

$$\vec{E}_2(\theta) = \begin{bmatrix} \cos\theta + \sin\theta e^{j\Delta\varphi} \\ 0 \end{bmatrix}. \qquad (11)$$

Taking into account (10) and (11), the signals at the inputs of a two-channel receiver will have the form

$$\dot{E}_1(\theta) = \frac{1}{\sqrt{2}}\{-\sin\theta + \cos\theta\cos\Delta\varphi + j\cos\theta\sin\Delta\varphi\}; \qquad (12)$$

$$\dot{E}_2(\theta) = \frac{1}{\sqrt{2}}\{\cos\theta + \sin\theta\cos\Delta\varphi + j\sin\theta\sin\Delta\varphi\}. \qquad (13)$$

Complex elements (12) and (13) are projections of the elliptically polarized resulting wave (3) on the LPS orts oriented at the angle $\theta$ to the measurement plane.

Let's find amplitudes $A_1$ and $A_2$, and phases $\Psi_1$ and $\Psi_2$ of the signals (12) and (13) at the output of the two-canal receiver with linear amplitude characteristic and linear detector and also find their relation with the bearing $\alpha$ and the angle $\theta$.

If the bearing is found using the amplitudes $A_1$ and $A_2$ ratio the method is called polarization - amplitude, if the phase $\Psi_1$ and $\Psi_2$ difference is used the method is called polarization - phase.

**Polarization-amplitude method of a moving object bearing**

Amplitudes $A_1$ and $A_2$ of the signals (12), (13) can be written

$$A_1(\theta) = \frac{1}{\sqrt{2}}\sqrt{1-\sin 2\theta \cos\Delta\varphi}; \qquad (14)$$

and

$$A_2(\theta) = \frac{1}{\sqrt{2}}\sqrt{1+\sin 2\theta \cos\Delta\varphi}. \qquad (15)$$

Ratio of the receiver output signals

$$\frac{A_1(\theta)}{A_2(\theta)} = \frac{\sqrt{1-\sin 2\theta \cos\Delta\varphi}}{\sqrt{1+\sin 2\theta \cos\Delta\varphi}}. \qquad (16)$$

Ratio (16) shows dependence as on the MO angle co-ordinate as on the angle $\theta$ LPS.

After putting $\theta = \pi/4$ in (16)

$$\frac{A_1}{A_2} = \frac{\sqrt{1-\cos\Delta\varphi}}{\sqrt{1+\cos\Delta\varphi}} = |\mathrm{tg}\,\Delta\varphi/2|, \qquad (17)$$

So

$$\Delta\varphi = \pm 2\arctan\frac{A_1}{A_2} \pm n2\pi, \qquad (18)$$

where $n = 0, 1, 2...$

The MO bearing $\alpha$ can be calculated after putting (18) in (2)

$$\alpha = \arcsin\left[\frac{\lambda}{\pi d}\left(\pm\arctan\frac{A_1}{A_2} \pm n\pi\right)\right]. \qquad (19)$$

Equation (19) shows that in order to calculate the MO bearing $\alpha$ it is necessary and enough to measure the ratio of the amplitudes of the signals from the outputs of the LPS arms provided that $\theta = \pi/4$.

Dependence (19) is not univocal. The unicity zone width $\Delta\alpha$ in the direction about zero can be calculated by (19) when $n = 0$ and $A_1/A_2 = \infty$

$$\Delta\alpha = \arcsin\frac{\lambda}{2d}. \qquad (20)$$

The dependence of the ratio $A_1/A_2$ on the angular coordinate of the point is called the direction finding characteristic of the polarization-amplitude angle measuring system. The steepness of this characteristic is minimum at $\alpha = 0$. Using the relations (17) and (1), it can be shown that the steepness of the direction finding characteristic at point $\alpha = 0$ is defined as:

$$\frac{d\left(\frac{A_1}{A_2}\right)}{d\alpha} = \frac{d}{d\alpha}\left\{\mathrm{tg}\left(\frac{\pi d}{\lambda}\sin\alpha\right)\right\}\bigg|_{\alpha=0} = \frac{\pi d}{\lambda}. \qquad (21)$$

From the practical point of view the zone of the any single-valued measurement of any angle measuring system should as wide as possible and the bearing characteristic steepness maximal. Analysis of (20) and (21) shows that in case of polarization-amplitude system these demands are contradictory because in order to wide the zone it is necessary minimize the base $d$. Solution of this contradiction promise multi-base polarization-amplitude systems.

**Polarization-phase method of the moving object bearing**

Phases $\Psi_1$ $\Psi_2$ of the signals (12) (13) are:

$$\Psi_1(\theta) = \mathrm{arctg}\frac{\cos\theta\sin\Delta\varphi}{-\sin\theta+\cos\theta\cos\Delta\varphi}; \quad (22)$$

$$\Psi_2(\theta) = \mathrm{arctg}\frac{\sin\theta\sin\Delta\varphi}{\cos\theta+\sin\theta\cos\Delta\varphi}. \quad (23)$$

The phase difference

$$\Delta\Psi(\theta) = \Psi_1(\theta) - \Psi_2(\theta) = \mathrm{arctg}\left[\frac{1}{\cos 2\theta}\mathrm{tg}\,\Delta\varphi\right]. \quad (24)$$

In general case the phase difference $\Delta\Psi(\theta)$ of the signals (12) and (13) depends not only on the MO angle co-ordinate but also on the LPS orientation $\theta$ to the measurement plane.

When in (24) $\theta = \pi/4$, the phase difference is constant, equal $\pi/2$ and does not depend on the MO angle co-ordinate. In such a case, as follows from (17), the MO angle co-ordinate depends only on the ratio of the signal amplitudes LPS shoulder outputs.

If $\theta = 0°$ in (24), $\Delta\Psi$ at the LPS output depend only on the phase difference $\Delta\varphi$ of the orthogonal linearly polarized waves at the receiving point at the input of the receiving onboard antenna

$$\Delta\Psi = \Delta\varphi. \quad (25)$$

The amplitudes ratio is constant, equal 1 ($A_1/A_2 = 1$,) and does not depend on the MO angle co-ordinate $\alpha$.

After putting (25) in (2)

$$\alpha = \pm\arcsin\frac{\lambda}{2\pi d}(\Delta\Psi) \pm n\pi, \quad (26)$$

where $n = 0, 1, 2...$.

Equation (26) shows that in order to calculate the MO bearing it is necessary to measure the signals (12) and (13) phase difference $\Delta\Psi$ at the outputs of the LPS with angle orientation $\theta = 0°$.

As the measuring device there is usually used phase detector [3,4]. As phase measuring systems provide univocal result within some interval $\pm\varphi_0$, (26) gives univocal result $\alpha$ in the sector

$$\pm\alpha_0 = \arcsin\left(\pm\frac{\varphi_0\lambda}{2\pi d}\right). \quad (27)$$

At the output of the phase measuring device, a voltage $U_{pd}$ is formed. The sign and magnitude of this voltage in a certain sector of angles makes it possible to unambiguously determine the bearing $\alpha$ as [6]

$$U_{pd} = k\sin\Delta\varphi, \quad (28)$$

where $k$ is the coefficient depending on phase detector parameters and amplitudes of the signals which phase difference is measured. Usually there is used automatic gain control and signal limiting to make the amplitude of the signal at the phase detector input constant. Putting (1) in (28), we obtain a discriminatory characteristic:

$$U_{pd}(\alpha) = U_0 \sin\left(\frac{2\pi d}{\lambda}\sin\alpha\right), \quad (29)$$

where $U_0 = const$.

If $\alpha$ are not big the dependence (29) is proximally linear

$$U_{pd}(\alpha) = U_0 \frac{2\pi d}{\lambda}\sin\alpha. \quad (30)$$

One can see (30) that the value and the sign of $\alpha$ can be calculated if the voltage at the phase detector is known. Dependence of $U_{pd}/U_0$ on $\alpha$ is called direction finding characteristic of an angle measuring system

$$F(\alpha) = \frac{U_{\phi.\partial}(\alpha)}{U_0} = \frac{2\pi d}{\lambda}\alpha. \quad (31)$$

Its derivative at $\alpha = 0$ is called the steepness of the direction finding characteristic or the sensitivity of the bearing.

$$S(\alpha) = \left|\frac{dF(\alpha)}{d\alpha}\right|_{\alpha=0} = \frac{2\pi d}{\lambda}. \quad (32)$$

So, sensitivity, and as the result accuracy of bearing, grow when grows the ratio $d/\lambda$. However, at $d/\lambda > 1/2$, there appears an ambiguity in measuring the angle, which follows from the expression (29). It could be removed by measuring at different $d/\lambda$ with several scales as in phase direction finder.

**Conclusion**

As the result of investigation we can conclude:

1. If the radio beacon illuminates simultaneously from two spaced points horizontally and vertically polarized signals with equal amplitudes, initial phases and wave lengths and the resulting vector signal are received onboard MO within LPB with orts coinciding the LPS own orts and oriented at $\theta = \pi/4$, then the MO bearing $\alpha$ is calculated from the ratio of the amplitudes of the signals from the outputs of the LPS arms.

2. If LPB orts coincide the LPS own orts and oriented at $\theta = 0°$, then the MO bearing is calculated using the measured phase difference from the outputs of the LPS arms.

3. The steepness of the direction finding characteristic, and hence the potential accuracy of direction finding, is determined in both methods by the spatial separation $d$ of radiation sources, and not onboard receiving antenna pattern.

4. The bearing sensitivity at the directions about $\alpha = 0°$ twice higher in case of polarization - phase method then in case of polarization - amplitude one.

**Acknowledgments**
This study was supported by the Ministry of Science and Education of the Russian Federation, project no. FEWM-2020-0039.